\documentclass[a4paper,letters,fleqn,usenatbib]{mnras}

\usepackage{newtxtext,newtxmath}

\usepackage[T1]{fontenc}
\usepackage{ae,aecompl}

\usepackage{graphicx}	
\usepackage{amsmath}	
\usepackage{amssymb}	

\usepackage[dvipsnames]{xcolor}
\usepackage{units}
\usepackage{textcomp}
\usepackage{colortbl}

\title[First \textit{L}~band detection of hot exozodiacal dust with VLTI/MATISSE]{\vspace*{-1cm}First \textit{L}~band detection of hot exozodiacal dust with VLTI/MATISSE\\[-0.7cm]}

\author[F. Kirchschlager et al.
]{
F. Kirchschlager$^{1}$\thanks{\vspace*{-0.4cm}E-mail: \href{mailto:f.kirchschlager@ucl.ac.uk}{f.kirchschlager@ucl.ac.uk}}, S. Ertel$^{2,3}$, S. Wolf$^{4}$, A. Matter$^{5}$ and A.~V.~Krivov$^{6}$\\
$^{1}$Department of Physics and Astronomy, University College London, Gower Street, London WC1E 6BT, United Kingdom\\
$^{2}$Large Binocular Telescope Observatory, 933 North Cherry Avenue, Tucson, AZ 85721, USA\\
$^{3}$Steward Observatory, Department of Astronomy, University of Arizona, 993 N. Cherry Ave, Tucson, AZ, 85721, USA\\
$^{4}$Institute of Theoretical Physics and Astrophysics, Kiel University, Leibnizstra\ss e 15, 24118, Kiel, Germany\\
$^{5}$Laboratoire Lagrange, Universit\'e C\^ote d'Azur, Observatoire de la C\^ote d'Azur, CNRS, Boulevard de l'Observatoire, CS~34229,\\\phantom{$^{4}$ } 06304 Nice Cedex 4, France\\
$^{6}$Astrophysical Institute and University Observatory, Friedrich Schiller University Jena,  Schillerg\"a\ss chen 2-3, 07745 Jena, Germany\vspace*{-0.8cm}
}

\date{Accepted 2020 September 4. Received 2020 September 4; in original form 2020 July 14\vspace*{-0.6cm}}

\pubyear{2020}

\begin{document}
\label{firstpage}
\pagerange{\pageref{firstpage}--\pageref{lastpage}}
\maketitle

\begin{abstract}
For the first time we observed the emission of hot exozodiacal dust in \textit{L}~band. We used the new instrument MATISSE at the Very Large Telescope Interferometer to detect the hot dust around $\kappa$~Tuc with a significance of $3\sigma$ to $6\sigma$ at wavelengths between 3.37 and $\unit[3.85]{\mu m}$ and a dust-to-star flux ratio of $5$ to $\unit[7]{\%}$.
 We modelled the spectral energy distribution based on the new \textit{L}~band  data alone and in combination with \textit{H} band data published previously. In all cases we find $\unit[0.58]{\mu m}$ grains of amorphous carbon to fit the $\kappa$~Tuc observations the best, however, also nanometre or micrometre grains and other carbons or silicates reproduce the observations well. Since the \textit{H}~band data revealed a temporal variability, while our \textit{L}~band data were taken at a different epoch, we combine them in different ways. Depending on the approach, the best fits are obtained for a narrow dust ring at a stellar distance in the $0.1$ to $\unit[0.29]{au}$ range and thus with a temperature between
$940$ and  $\unit[1430]{K}$.
Within the $1\sigma$ uncertainty dust location and temperature are confined to $\unit[0.032 - 1.18]{au}$ and $\unit[600 - 2000]{K}$. 
\end{abstract}

\begin{keywords}
\vspace*{-0.1cm}zodiacal dust -- infrared: planetary systems -- techniques: interferometric\vspace*{-0.75cm}
\end{keywords}

 

\section{Introduction}
\label{101}
Since the first detection of hot exozodiacal dust (``hot exozodi'') around Vega (\citealt{Absil2006}), about two dozens hot exozodis have been discovered  using optical long baseline interferometry (\citealt{Absil2008, Absil2009, Absil2013, DiFolco2007, Defrere2011, Defrere2012,  Ertel2014, Ertel2016, Nunez2017}). Presumably accumulating at or close to the sublimation radius, the dust is heated to high temperatures and its emission peaks in the near- (NIR) to mid-infrared (MIR). The detected dust-to-star flux ratios in NIR are at a level of a few per cent or even lower (e.g. \citealt{Ertel2014}). Thus, high precision (contrast) and high angular resolution (${\sim}\unit[0.01]{as}$) are required to observe hot exozodis.

The existence of hot exozodis raises questions as dust located at stellar distances of only ${\sim}\unit[0.01-1]{au}$  would be removed by radiative forces on timescales of a few years. To be\break detectable around ${\sim}20\,\%$ of main-sequence stars of all spectral types from A to K at all ages (\citealt{Ertel2014}), the dust has to be continuously replenished or to be trapped in the stellar vicinity for long times, yet the eliciting mechanism has still to be identified (\citealt{vanLieshout2014, Rieke2016, Kral2017, Kimura2020, Pearce2020}).\break\vspace*{-0.4cm} 

Studies of hot exozodis offer a way to better understand the inner regions of extrasolar planetary systems. In addition, exozodis could help to trace the invisible sources of the dust and the putative planets at larger distances, and hence reveal the architecture of the planetary systems. On the other hand, the possible presence of small grains is a potential problem for the detection of terrestrial planets in the habitable zone (e.g. \citealt{Agol2007, Beckwith2008}).

So far, hot exozodis have only been observed in \textit{H} or \textit{K}~band, thus grain sizes and grain compositions could not be constrained sufficiently. Previous modelling of the Spectral Energy Distribution (SED) pointed towards nano- to \mbox{sub-}micrometre sized grains of carbonaceous material, however, larger grains could not be ruled out completely (\citealt{Kirchschlager2017}). \textit{N}~band emission from more temperate (warm) dust near habitable zones has been detected (\citealt{MillanGabet2011, Mennesson2014, Ertel2018, Ertel2020}) but observations at intermediate wavelengths ($\unit[1.6]{\mu m}\lesssim\lambda\lesssim\unit[10]{\mu m}$) are required to study the potential connection of warm and hot dust.

The Multi AperTure mid-Infrared SpectroScopic Experiment (MATISSE; \citealt{Lopez2014}) is a second-generation instrument at the Very Large Telescope Interferometer (VLTI), available since 2019. With a spatial resolution of a\break few mas and operating in \textit{L}, \textit{M}~and \textit{N}~band, MATISSE offers\break critical capabilities for the study of hot exozodis. In particular, MATISSE will be able to confine the dust properties of hot exozodis (\citealt{Ertel2018b, Kirchschlager2018}). 

$\kappa~$Tuc (HD~7788) is an \mbox{F6 IV-V} star located in the constellation Tucana at a distance of $\unit[(21.0\pm0.3)]{pc}$ (\citealt{Gaia2018}) with an effective temperature of $\unit[6474]{K}$, stellar mass ${\sim}\unit[1.35]{M_\odot}$, and an age of ${\sim}\unit[2]{Gyr}$ (\citealt{Ammler2012, Fuhrmann2017, Tokovinin2020}). Significant hot emission was detected in 2012 and 2014 around $\kappa~$Tuc (\citealt{Ertel2014}). However, no significant excess was detected in 2013, making $\kappa~$Tuc the first hot exozodi candidate for significant NIR variability (\citealt{Ertel2016}).\break\vspace*{-0.4cm} 

In this letter, we present observations in \textit{L}~band of the hot exozodi around $\kappa$~Tuc (Section~\ref{sec_obs}) and a modelling of the observed visibilities and SED with the focus to constrain the dust properties in the circumstellar environment (Sections~\ref{sec3} and \ref{sec_results}). We discuss the results in Section~\ref{sec_conc}.

\begin{figure*}
\includegraphics[trim=1.6cm 1.4cm  3.4cm 2.3cm,  clip=true, width=0.3582\linewidth, page = 1]{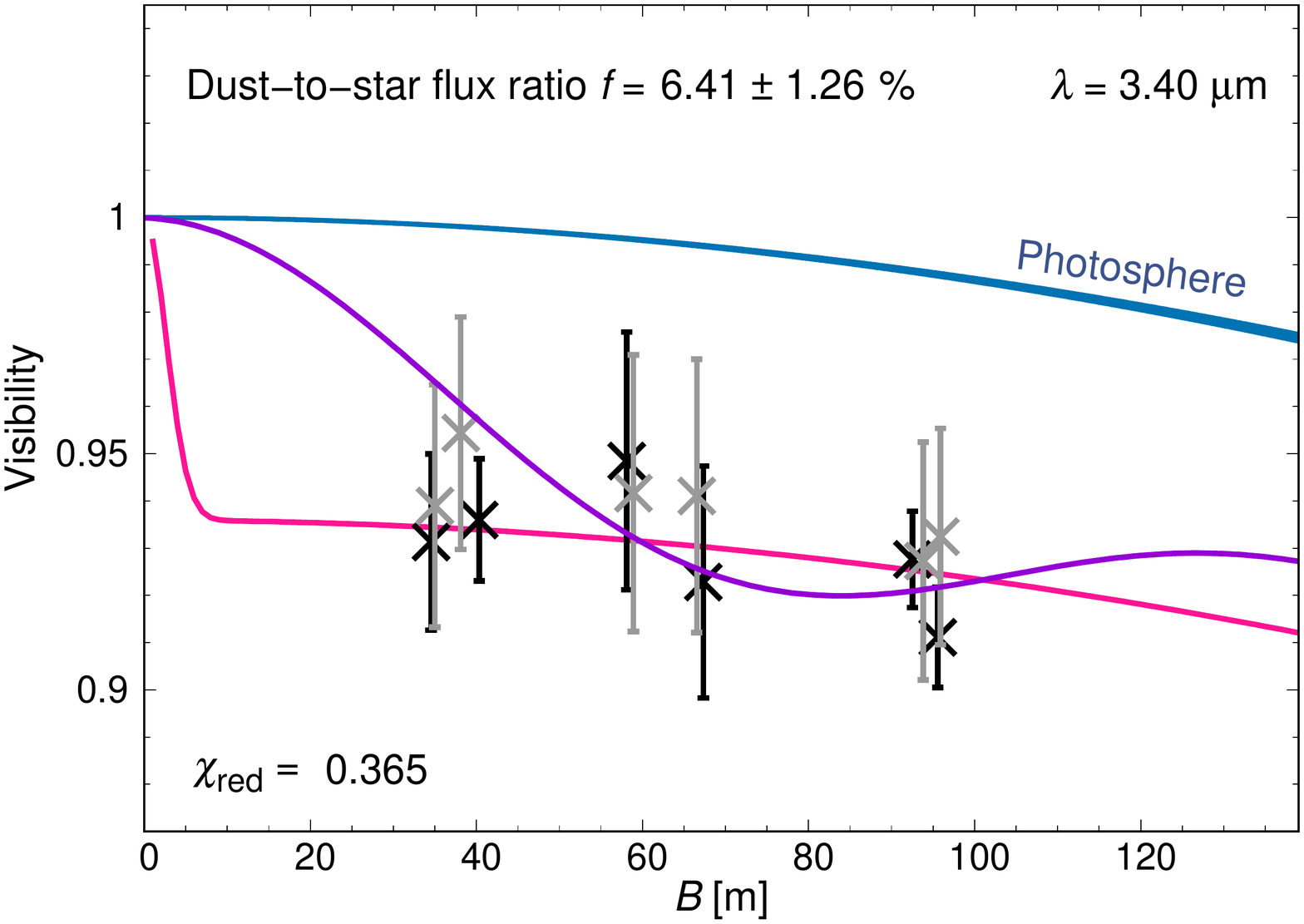}\hspace*{-0.08cm}  
\includegraphics[trim=4.2cm 1.4cm  3.4cm 2.3cm,  clip=true, width=0.32\linewidth, page = 2]{Pics/Visibility_HD7788_Paper_referee2.pdf}\hspace*{-0.08cm} 
\includegraphics[trim=1.6cm 1.4cm  5.9cm 2.3cm,  clip=true, width=0.3225\linewidth, page = 1]{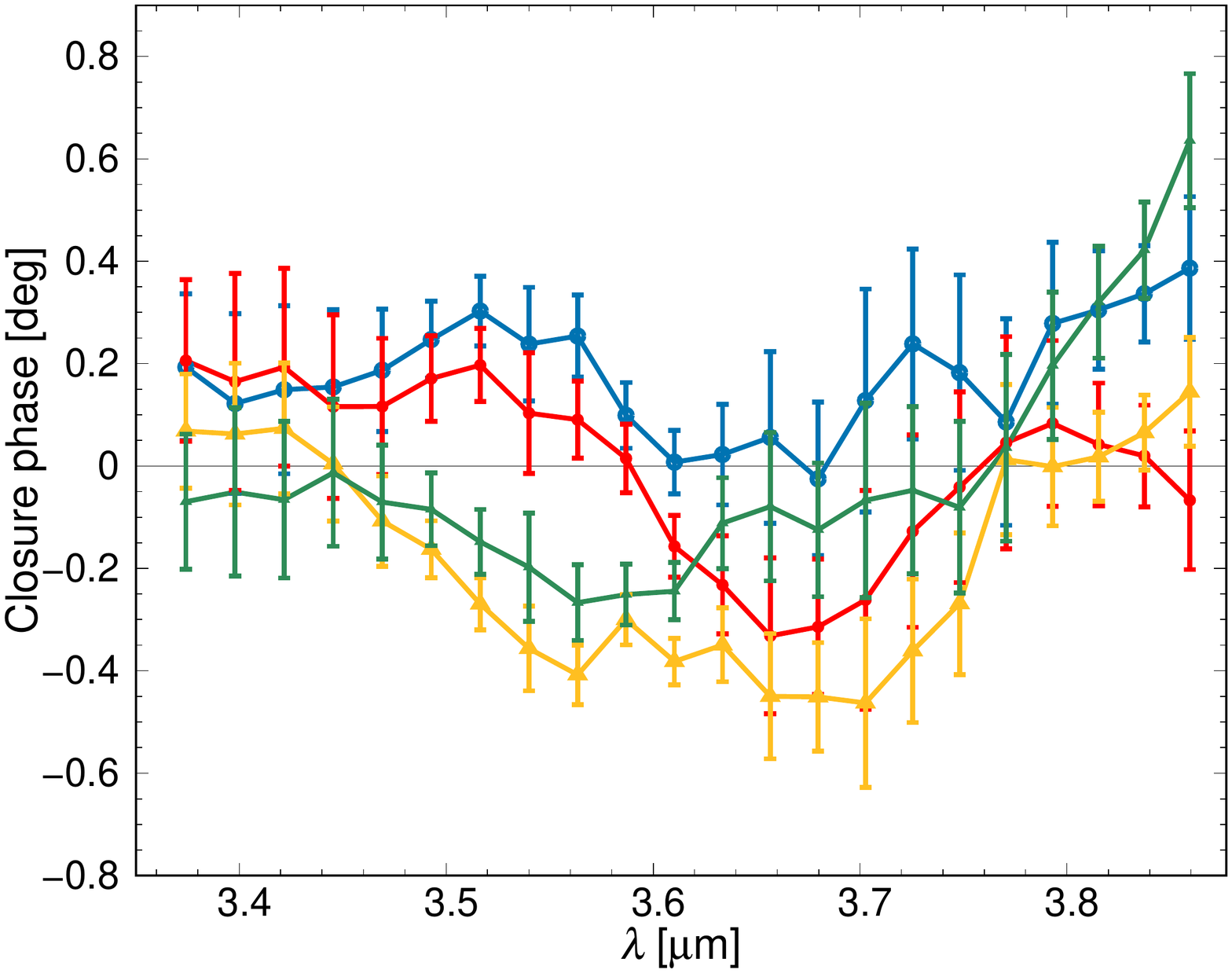}\\[-0.2cm]
\caption{\textit{Left \& Centre}: Measured visibilities and related 1$\sigma$~errors (crosses and bars; observations on 9 July 2019 in black and on 11 July 2019 in grey) along with the expected visibility of the limb-darkened photosphere (blue line) as a function of projected baseline length $B$, for the wavelengths  $\lambda=\unit[3.4]{\mu m}$ and $\unit[3.61]{\mu m}$. The thickness of the blue line corresponds to an adopted uncertainty of the stellar diameter. The best-fit model is represented by the magenta line and corresponds to a uniform circumstellar emission (Gaussian profile) with a wavelength-dependent dust-to-star flux ratio $f$. The purple line represents the best-fit of a disc modelling approach (Section~\ref{sec_results}). \textit{Right}: Calibrated closure phases for the four triangles of the observation on 9 July. The 22 spectral elements range from $\unit[3.37]{\mu m}$ to $\unit[3.85]{\mu m}$ and cover the region that show significant dust emission (see Section~\ref{sec3} for further details).\vspace*{-0.3cm}}                                      
\label{fig_visdata} 
\end{figure*}

\section{Observations}
\label{sec_obs}
Interferometric \textit{LM}~and \textit{N}~band data were obtained on 9 and 11 July 2019 using the instrument MATISSE (\citealt{Lopez2014}) on the VLTI (Table~\ref{tab_obs}). The Auxiliary Telescopes (ATs) were arranged in medium configuration and the New Adaptive Optics Module for Interferometry (NAOMI; \citealt{Woillez2019}) was used. We simultaneously obtained visibility measurements on six baselines (baseline lengths between $B=\unit[30]{m}$ and $\unit[95]{m}$). The observations were carried out in LOW spectral resolution (R$\sim$30). In \textit{LM}~band, the fringes were dispersed over 64 spectral pixels between $3.28$ and $\unit[4.57]{\mu m}$, which correspond to about 13 true spectral channels (spectral channels are sampled over 5 spectral pixels on the \textit{LM}~band detector). Our measurements thus cover a significant part of the \textit{L}~band and the very beginning of the \textit{M}~band. 
In \textit{N} band, $\kappa$~Tuc ($N\sim\unit[1.5]{Jy}$) is too faint for visibility measurements with MATISSE (to date sensitivity limit ${\sim}\unit[15-20]{Jy}$; see MATISSE ESO webpage). Even though \textit{N} band interferometric data were acquired simultaneously to the \textit{LM}~band data, the independent step of \textit{N}~band photometric measurements was thus skipped. No \textit{N} band visibility could be computed and we discarded the \textit{N} band data. In this work, our focus is on the \textit{LM} band data only.\break\vspace*{-0.35cm} 

Observations$\,$of$\,\kappa$~Tuc$\,$were$\,$framed$\,$by$\,$observations$\,$of$\,$two reference stars to calibrate the instrumental contribution (CAL-SCI-CAL sequence per night).$\,$Calibrators$\,$were$\,$chosen\break from the SearchCal tool (\citealt{Chelli2016}) to be regular main-sequence stars similar in magnitude and position of $\kappa~$Tuc and with a stellar angular diameter as low as possible.\break\vspace*{-0.4cm} 


The MATISSE data were reduced and calibrated with the$\,$help$\,$of~SUV,$\,$the$\,$VLTI$\,$user$\,$support$\,$service$\,$of$\,$the$\,$JMMC\footnote{Website: \href{http://www.jmmc.fr/suv.htm}{http://www.jmmc.fr/suv.htm}. Here, the last version of the MATISSE pipeline was used which is publicly available at \href{http://www.eso.org/sci/software/pipelines/matisse/}{http://www.eso.org/sci/software/pipelines/matisse/}.\\[-1.3cm]}. A specific aspect of MATISSE is the presence of two Beam Commuting Devices (BCDs) that provide four independent\break beam configurations to calibrate$\,$out$\,$the$\,$effect$\,$of$\,$instrumental defects (\citealt{Lopez2014}): IN-IN, IN-OUT, OUT-IN, and OUT-OUT. A basic observation cycle in \textit{LM}~band thus consists of four 1-min interferometric + photometric exposures, each of them being associated with one BCD position. Here, we focus on the absolute visibilities $V$ which requires to properly examine the four exposures for their visibility accuracy. For the observation on 9~July  it turned out that the OUT-OUT exposure had to be discarded as the corresponding visibilities are inconsistent (within 3$\sigma$) with the visibilities of the three other exposures, for most of the baselines. For the same reason, we discarded the IN-IN exposure of the 11 July observation.

  \begin{table}
 \vspace*{-0.3cm}
 \centering
 \caption{Observations of $\kappa$~Tuc (HD~7788) with VLTI/MATISSE (Program No. 0103.C-0725(A); PI: F. Kirchschlager)\vspace*{-0.15cm}}
 \begin{tabular}{ l l l l l}
 \hline
ID & Date	& Config. 	& Seeing 	&Calib.${}^{\star}$\\\hline
A  & 2019/07/09	& K0-G2-D0-J3	& $\unit[0.8]{as}$	& 1, 2\\
B  & 2019/07/11	& K0-G2-D0-J3	& $\unit[0.8]{as}$	& 3, 4\\
\hline
  \multicolumn{5}{l}{\textbf{Notes.} $({}^{\star})$ Calibrator stars correspond to}\\
  \multicolumn{5}{l}{(1) HD 3750 (K1III); (2) HD 8094 (K4III);}\\
  \multicolumn{5}{l}{(3) HD 4138 (K4III); and (4) HD 8315 (K0III).}\\[-0.3cm]
 \end{tabular}
 \label{tab_obs}
 \end{table}

Besides the `short-term' errors of the pure fundamental  source noise, thermal background and detector readout noise, broadband errors on the photometry arise due to variations of the interferometric transfer function as well as due to an imperfect subtraction of the thermal background, which  affect the visibility on a timescale of $\sim$min. The transfer function shows variations of less than 2 per cent on average, which is thus not a limiting factor for the visibility accuracy. We take the mean of the three exposures as final visibility for each spectral pixel and their `long-term' error is estimated by the standard deviation of the three exposures.

Hence, the data set comprises $2\,\text{(observations)}\,\times\,6\,\text{(baselines)}$ = 12 independent visibility measurements for each wavelength. The final calibrated data set of two exemplary wavelengths plus the closure phase are shown in Fig.~\ref{fig_visdata}.\break\vspace*{-0.7cm}

\section{Analysis of the MATISSE data}
\label{sec3}
When we compare the calibrated data to the expected visibility of the stellar photosphere a visibility deficit is revealed. A stellar companion can be rejected as a source of this deficit as the closure phases are close to zero (Fig.~\ref{fig_visdata}), in agreement with measurements by \cite{Marion2014}. Therefore, the visibility deficit must be caused by circumstellar emission in the field-of-view (FOV). We follow a two-step approach to determine the excess and the properties of the circumstellar  material. In the first step (this section), the calibrated visibilities $V$ are fitted by a model consisting of a limb-darkened photosphere surrounded by a uniform disc emission filling the entire FOV of \mbox{MATISSE}. In the second step (Section~\ref{sec_results}), the fluxes derived in the first step are fitted to a disc model where the dust is arranged in a narrow ring and the grains' optical properties are considered. The two-step approach has been used in previous studies to constrain the dust properties (\citealt{Absil2006, Absil2009, DiFolco2007, Absil2008, Defrere2011, Lebreton2013, Kirchschlager2017}) and we will demonstrate at the end of Section~\ref{sec_results} that the visibilities of the disc model are consistent  to the MATISSE data which justifies this approach.

 Using the flux ratio $f$ between the integrated circumstellar and the stellar photospheric emission, the combined visibility with contributions from the bare photosphere and from the circumstellar emission is (\citealt{DiFolco2007})
 \begin{align}
 V_{(\star+\text{CSE})} (B)&=\,(1-f)\,V_\star(B) + f\,V_\text{CSE}(B),\,\text{where}\\
   V_\star(B) &=  \frac{6}{3-u_\lambda} \left(\left(1-u_\lambda\right)\frac{J_1(x)}{x} + u_\lambda \sqrt{\frac{\pi}{2}} \frac{J_{1.5}(x)}{x^{1.5}}\right) \label{eq1}\\
   \text{and}\hspace*{0.2cm}V_\text{CSE}(B) &= \exp{\left(-\frac{x^2_\text{FOV}}{4 \ln{2}}\right)} \label{eq2}
 \end{align} 
are the visibilities of a limb-darkening stellar model (\citealt{HanburyBrown1974}) and of the circumstellar emission (symmetric Gaussian).
Here, $x =\pi \Theta_\star B/\lambda$, $\Theta_\star=\unit[(0.739\pm0.011)]{mas}$ is the stellar diameter (\citealt{Ertel2014}),  $u_\lambda=0.22$ is the linear limb-darkening coefficient in \textit{LM}~band (\citealt{Claret1995}), $J_1(x)$ and $J_{1.5}(x)$ are Bessel functions, $x_\text{FOV} =\pi \Theta_\text{FOV}B/\lambda$, and $\Theta_\text{FOV}=\unit[0.6]{as}$ is the physical FOV in \textit{LM}~band which is equivalent to $\unit[6.3]{au}$ in radius at the distance of $\kappa$~Tuc. Under typical seeing conditions, the sensitivity along the FOV follows a Gaussian profile with a full width at half maximum of the size of the FOV, FWHM$= \unit[0.6]{as}$.

The dust-to-star flux ratio $f$ is the only free parameter of our model which has to be fitted for each wavelength. To find the best-fit model, the calibrated MATISSE data of both observation nights are combined and fitted together (12 visibilities for each wavelength), the reduced $\chi^2$ is calculated as a function of $f$ and minimised. The statistical errors of $f$ are obtained by evaluating the probability distribution $p\left(\chi^2\right)=\exp{\left(-\chi^2 -\chi^2_\text{best-fit}\right)}$ around the best-fit model and by determining the corresponding confidence levels of $p$.  

The visibility distribution of the best-fit model has been overlaid on the plots of the calibrated data (Fig.~\ref{fig_visdata}, magenta line). The results of the fitting procedure are displayed as insets together with the 1$\sigma$ error of the flux ratio, $\pm \Delta f$, and the\break reduced $\chi^2$. The detection of the flux is significant when $f/|\Delta f|\ge3$ which accounts for 22 wavelengths between 3.37 and $\unit[3.85]{\mu m}$ which cover only the \textit{L}~band.  This is the first time ever hot exozodiacal dust has been detected in this waveband. The flux ratios of  about $\unit[5-7]{\%}$ are distinctly higher than the ratios known from \textit{H}~and \textit{K}~band for typical\break hot exozodis ($\unit[{\sim}1]{\%}$; \citealt{Absil2013,Ertel2014}). On the other hand, no significant (${<}3\sigma$) emission is detected at\break wavelengths above $\unit[3.85]{\mu m}$. This is not because of larger uncertainties of the individual exposures but due to an increased dispersion of the data points of the exposures. The visibilities at  wavelengths below $\unit[3.37]{\mu m}$ increase to larger values and show either no significant$\,$deficit$\,$or$\,$no$\,$deficit$\,$at$\,$all.

\begin{figure}
\centering                                       
\includegraphics[trim=1.7cm 1.35cm 3.4cm 5.9cm,  clip=true, width=0.9\linewidth, page = 1]{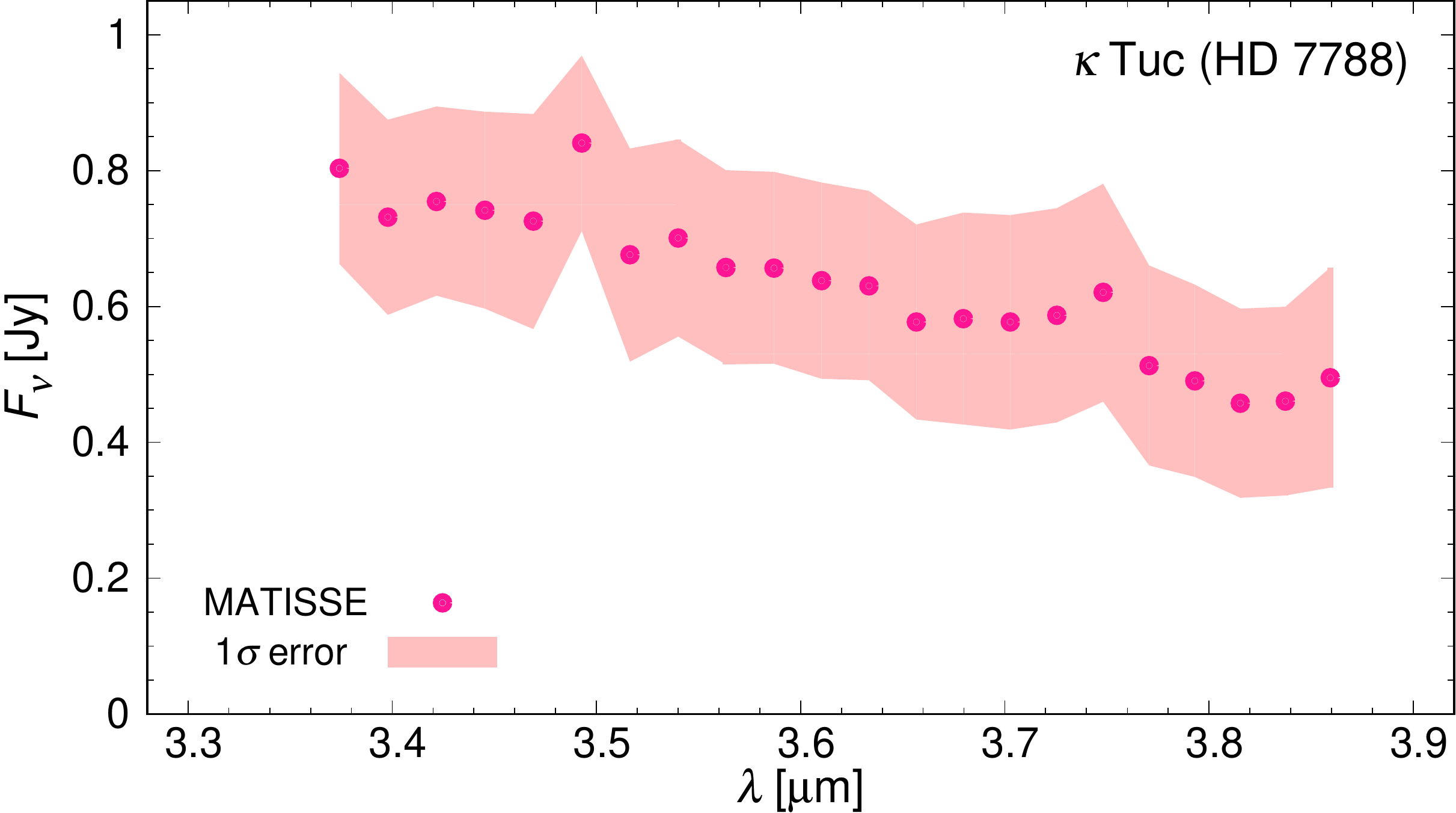}\\[-0.1cm] 
\caption{Spectral energy distribution (SED) of the circumstellar dust inferred from \textit{L}~band observations with ${>}3\sigma$ significance.\vspace*{-0.3cm}}                                      
\label{fig_SED} 
\end{figure}

Considering the emission of a stellar black-body ($T_\text{eff}=\unit[6474]{K}$) as a weighting factor for $f$, we calculated the flux of the significant circumstellar emission (Fig.~\ref{fig_SED}). The fluxes show a steeply decreasing slope $F_\nu\propto \lambda^{-\alpha}$ with $\alpha= 3.92_{-1.92}^{+2.68}$. Two features are visible in the spectrum at $\unit[3.5]{\mu m}$ and $\unit[3.75]{\mu m}$. However, these features are not significant as can be seen by distinctly higher values for the reduced $\chi^2$ of $2.491$ for  $\lambda =\unit[3.5]{\mu m}$ and $2.306$ for  $\lambda =\unit[3.75]{\mu m}$ while it is about $1-1.5$ for neighbouring wavelengths, and are produced by a single visibility measurement on 11 July.\vspace*{-0.3cm} 

\section{Modelling of dust properties}
\label{sec_results}
Besides the newly obtained MATISSE data, the hot exozodi of $\kappa$~Tuc has been observed three times in \textit{H}~band using VLTI/PIONIER (\citealt{Ertel2014, Ertel2016}). In 2012 and 2014, the circumstellar emission was significant with a flux ratio of  $f=\unit[(1.43\pm0.17)]{\%}$ and $\unit[(1.16\pm0.18)]{\%}$, respectively, while the emission in 2013 was insignificant with $f=\unit[(0.07\pm0.16)]{\%}$. 

A blackbody fit to the SED is found wanting and we can exclude once for all a stellar companion as source of the circumstellar emission. Instead, we fit the SED by a disc model relying on the approach of \cite{Kirchschlager2017}. In order to avoid parameter degeneracies, we keep the disk model as simple as possible and the number of free parameters as low as possible. Similar to the ring model of \cite{Absil2009} and \cite{Defrere2011}, our model is represented by a narrow face-on ring with $R$ as inner radius, outer radius $1.1\,R$, dust mass $M_\text{dust}$, and number density $n(r)\propto r^{-1}$. The ring is composed of compact spherical dust grains with radius $a$. Spherical grains with $2\,a\lesssim \lambda \lesssim 10\,a$ show in general 
strong interferences in the emission SED which can affect the fitting result. In order to reduce this effect we use a narrow size distribution with width $\Delta a =0.3\,a$ around $a$, where the presence of different sizes causes the interferences to cancel each other out (e.g. \citealt{Kirchschlager2019}). The radial distance is varied in the range $R\in[\unit[0.03]{au}, \unit[3]{au}]$, the grain size in the range 
$a\in  [\unit[1]{nm}, \unit[5]{\mu m}]$, the dust mass in the range $M_\text{dust}\in[\unit[10^{-15}]{M_\oplus},\unit[10^{-3}]{M_\oplus}]$, and three different carbons and one silicate material are considered: Amorphous carbon (\citealt{Rouleau1991}), amorphous carbonaceous dust analogues (\citealt{Jager1998}), crystalline graphite (\citealt{DraineLee1984, DraineLaor1993}), and  astronomical silicate (\citealt{WeingartnerDraine2001}).

The grains' optical properties are calculated on the basis of Mie theory. Maps of single-scattering and re-emission are generated using an enhanced version of the tool \textsc{debris} (\citealt{Ertel2011}). Dust located within an inner working angle $\Theta=\lambda / (\unit[4\times100]{m})$ is assumed to be unresolved and its radiation is removed from the maps. The heterogeneous sensitivity along the FOV is taken into account by multiplying the simulated maps with a Gaussian function with a FWHM of $\unit[0.4]{as}$ and $\unit[0.6]{as}$ for \mbox{PIONIER} and MATISSE, respectively. Finally, the SED is calculated from the synthesised maps and fitted to the observational data by minimising the reduced $\chi^2$ and by varying the four free parameters $R$, $a$, $M_\text{dust}$, and the dust material. The corresponding confidence levels are again calculated by evaluating the probability function $p\left(\chi^2\right)=\exp{\left(-\chi^2 -\chi^2_\text{best-fit}\right)}$ around the best-fit model.

\begin{figure}
\centering                                       
\includegraphics[trim=1.75cm 1.35cm 3.4cm 5.9cm,  clip=true, width=0.9\linewidth, page = 1]{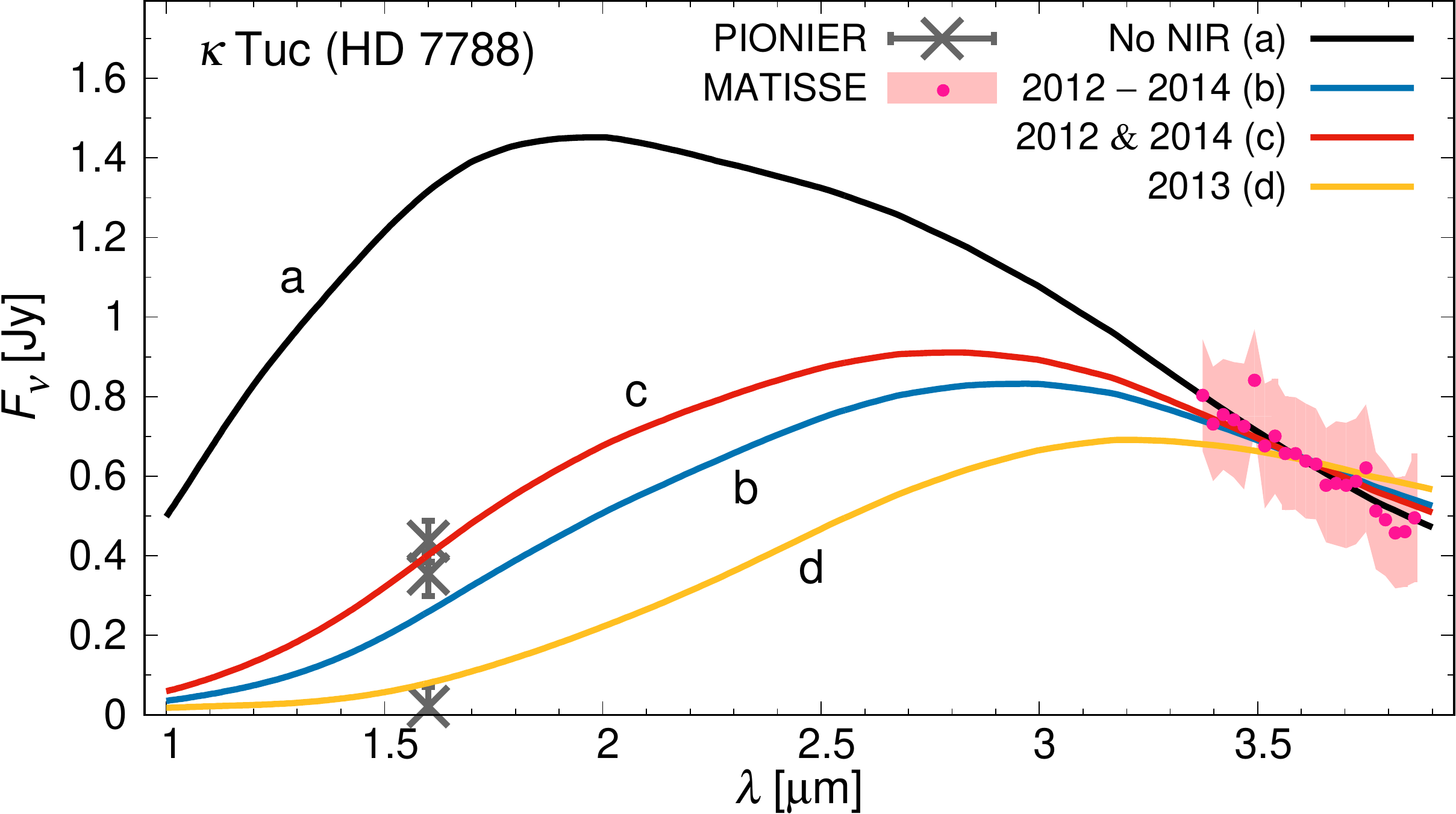}\\[-0.1cm]  
\caption{SED of the circumstellar dust emission around $\kappa$~Tuc composed of the \textit{H}~band (grey) observations  from 2012-2014 (\citealt{Ertel2014, Ertel2016}) and \textit{L}~band observations (magenta) from 2019 (this work). 
The solid lines represent the SEDs of the best-fit\break models for amorphous carbon taking into account four different fitting approaches: (a) Neglecting the NIR~fluxes (black), (b) considering all \textit{H}~band data (blue), (c) considering the 2012 and 2014 \textit{H}~band data (red), (d) considering 2013 \textit{H}~band data (yellow).\vspace*{-0.4cm}}                                      
\label{fig_bestSED} 
\end{figure}

\begin{figure}
\centering    
\includegraphics[trim=1.8cm 3.6cm 3.0cm 1.9cm,  clip=true, width= 0.9\linewidth, page = 1]{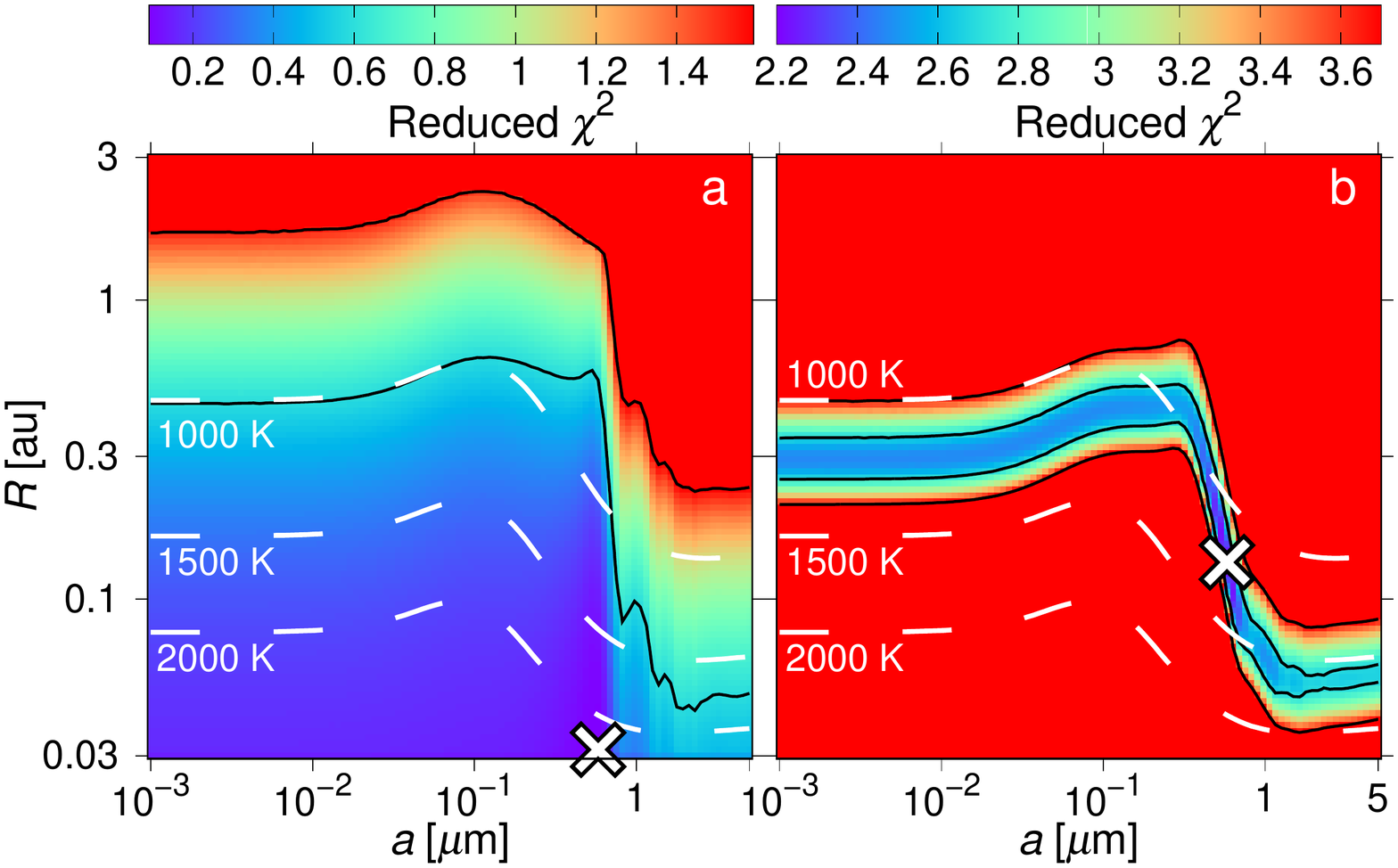}
\includegraphics[trim=1.8cm 3.6cm 3.0cm 1.9cm,  clip=true, width= 0.9\linewidth, page = 1]{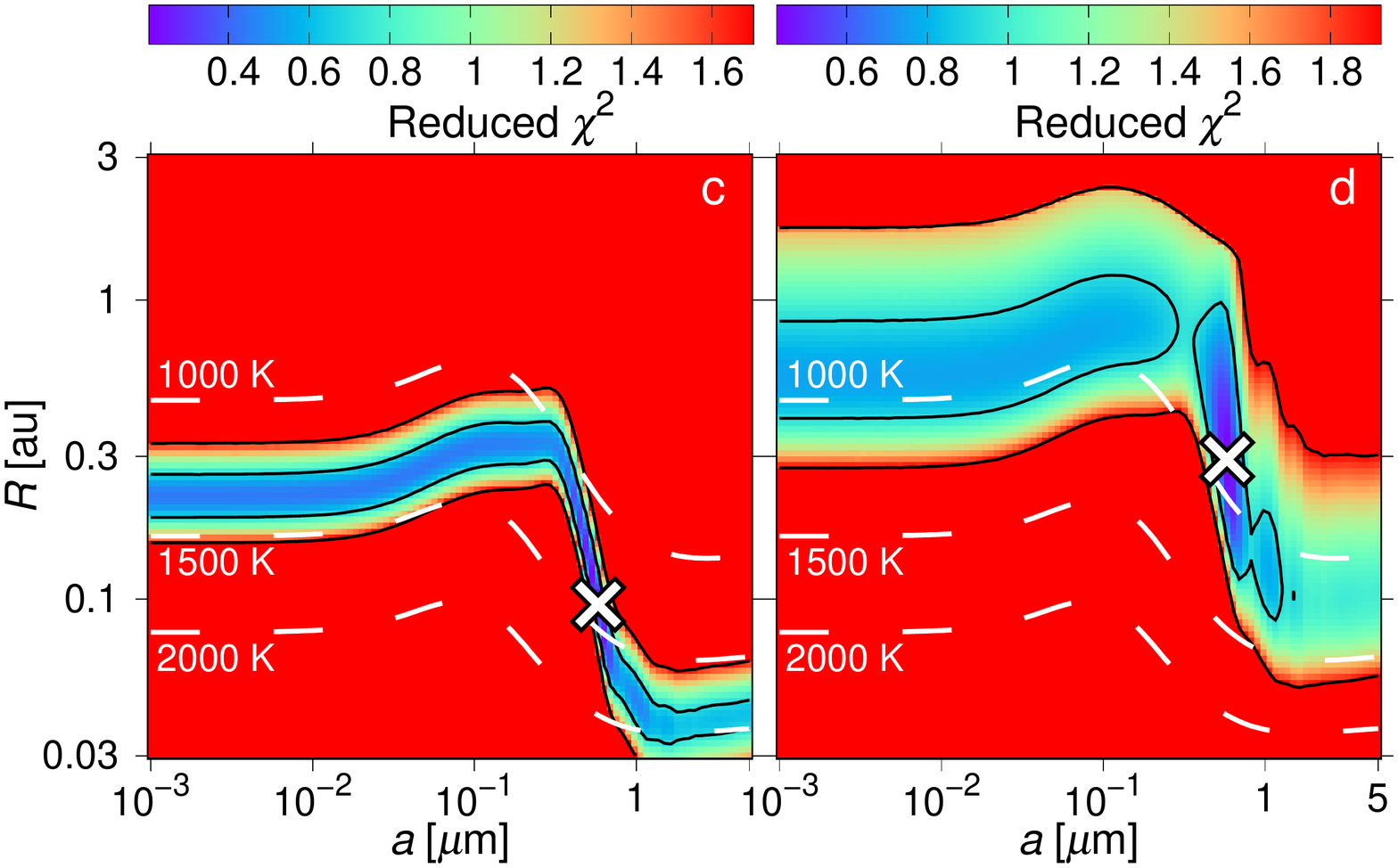}\\[-0.1cm] 
\caption{Reduced $\chi^2$ maps of the SED modelling of the hot exozodi of $\kappa$~Tuc  as  a function of grain size $a$ and disc radius $R$ for the four different fitting approaches (a) - (d).  The white crosses indicate the best-fit models and the black solid lines the $1\sigma$ and $3\sigma$ confidence levels. The dust temperatures $T_\text{dust}=\unit[1000]{K}$, $\unit[1500]{K}$, and $\unit[2000]{K}$ are shown as white dashed lines. The dust material is amorphous carbon.\vspace*{-0.3cm}}                                      
\label{fig_bestfit} 
\end{figure}

The main challenge is the \textit{H}~band variability and the time that passed between the PIONIER and MATISSE observations, and we follow different fitting approaches to comply with. Firstly (a), we solely fit the 22 significant \textit{L} band data points from MATISSE ($\lambda\in[\unit[3.37]{\mu m},\unit[3.85]{\mu m}$]). These data provide reliable information on the dust at the time of the MATISSE observations. Secondly, we assume that the three \textit{H} band measurements in 2012 to 2014 provide a probable range of the \textit{H} band excess during the MATISSE observations. We thus explore three scenarios when fitting MATISSE and PIONIER data together: (b) We include all three PIONIER measurements as this will give us the most likely result and a reasonable uncertainty on the dust parameters from the uncertainty of the PIONIER data. (c)~We include only the two significant \textit{H} band excesses from 2012 and 2014. This provides a result where the \textit{H} band excess is on the high side of the observed range. (d)~We include only the \textit{H} band non-detection from 2013, which gives the strongest upper limit available for the low side of \textit{H} band excesses observed. The results of our four fits are summarised in Table~\ref{tab_res} and illustrated in Fig.~\ref{fig_bestSED} and~\ref{fig_bestfit}. 

 \begin{table}
 \vspace*{-0.3cm}
 \centering
 \caption{Best fitting results derived from SED modelling.\vspace*{-0.25cm}}
 \begin{tabular}{ l l l l l}
 \hline
Appr.           &$\hspace*{-0.3cm}$ $a$ [$\mu$m]	&$\hspace*{-0.3cm}$  $R$ [au] 	&$\hspace*{-0.3cm}$  $T_\text{dust}$ [K] 	&$\hspace*{-0.3cm}$ $M_\text{dust}\,[10^{-9}\,\text{M}_\oplus]$\\\hline
(a) No NIR 	&$\hspace*{-0.3cm}$ 0.58		&$\hspace*{-0.3cm}$  0.031 		&$\hspace*{-0.3cm}$ 2300 			&$\phantom{1}$0.54 \\
(b) 2012-2014  	&$\hspace*{-0.3cm}$ 0.58		&$\hspace*{-0.3cm}$  0.13		&$\hspace*{-0.3cm}$ 1260  			&$\phantom{1}$3.11 \\
(c) 2012$\,$\&$\,$2014 	&$\hspace*{-0.3cm}$ 0.58		&$\hspace*{-0.3cm}$  0.1   		&$\hspace*{-0.3cm}$ 1430  			&$\phantom{1}$2.0\\
(d) 2013 	&$\hspace*{-0.3cm}$ 0.58		&$\hspace*{-0.3cm}$ 0.29		&$\hspace*{-0.3cm}$ $\phantom{1}$940		&10.4  \\
\hline\\[-0.7cm]
 \end{tabular}
 \label{tab_res}
 \end{table}

For the first approach (a), all considered dust materials are able to reproduce the MIR slope, however, the disc radius and the grain size are weakly constrained with large 1$\sigma$ errors. For amorphous carbon the best-fit model is composed of a disc radius $R=\unit[0.031]{au}$, grain size $a=\unit[0.58]{\mu m}$, mass $M_\text{dust}=\unit[0.54\times10^{-9}]{M_\oplus}$, and the temperature is $\unit[2300]{K}$.  

For the second approach (b), we find amorphous carbon to be the best-fit material, which is able to reproduce the NIR and MIR fluxes and the MIR slope. The radiation is dominated by thermal re-emission
 which agrees with a detected lack of scattered radiation in polarisation observations (\citealt{Marshall2016}). The best result ($\chi^2=2.199$) is obtained for the grain size $a=\unit[0.58]{\mu m}$, disc radius $R=\unit[0.13]{au}$, mass $M_\text{dust}=\unit[3.11\times10^{-9}]{M_\oplus}$, and the dust temperature amounts to $T_\text{dust}{\sim}\unit[1260]{K}$. However, within the 1$\sigma$ confidence level, the grain size is not constrained by our modelling and both nanometre-sized grains and those as large as a few micrometre can reproduce the observations. The inner disc radius is limited to $\unit[0.045]{au}<R<\unit[0.52]{au}$ within 1$\sigma$ confidence which corresponds to dust temperatures between $\unit[900]{K}$ and $\unit[1750]{K}$. Besides amorphous carbon, the three other materials (amorphous carbonaceous analogues, crystalline graphite, astronomical silicate) show fitting results which are only slightly weaker ($\chi^2=2.325-2.387$).

 For the third and fourth approach (c and d), amorphous carbon is adopted as dust material. The best-fit model of the third approach has a disc radius $R=\unit[0.1]{au}$, the dust temperature is $T_\text{dust}{\sim}\unit[1430]{K}$, and the dust mass is $M_\text{dust}=\break \unit[2.0\times10^{-9}]{M_\oplus}$, and that of the fourth approach is $R=\unit[0.29]{au}$,  $T_\text{dust}{\sim}\unit[940]{K}$, and $M_\text{dust}=\unit[10.4\times10^{-9}]{M_\oplus}$.  Within the $1\sigma$ confidence level the dust location and temperature of both approaches are limited to $\unit[0.032-1.18]{au}$ and $\unit[600-2000]{K}$. The grain size of both approaches amounts again to $a=\unit[0.58]{\mu m}$ but is not constrained within the 1$\sigma$ confidence error. The dust properties and the SED of approach (b) are, as expected, the intermediate values of approach (c) and (d). 

Finally, we have to verify that the disc model is compatible with the interferometric data. We take the maps of the best-fits of the four approaches and compute their interferometric signal. The closure phases of all disc models are zero as a result of the face-on orientation. In Fig.~\ref{fig_visdata} the computed visibilities of the disc model are shown as a function of baseline for approach (c). We can see that they approximate both the observational data and the visibilities of the uniform circumstellar emission. In particular, the visibility deficits of both circumstellar emission models (disc ring and Gaussian) compared to the pure stellar photosphere are similar. The main difference are the oscillations occurring for the disc model which are caused by the limited extension of the disc ring and which have been noticed also in the studies of \cite{Absil2009} and \cite{Kirchschlager2018}. In summary, we conclude that the disc model is in line with the interferometric data.\vspace*{-0.6cm}

 \section{Discussion and Conclusions}
 \label{sec_conc}
 In this letter, we presented the first detection of hot exozodiacal dust emission in \textit{L}~band. We used the new instrument MATISSE at the VLTI to observe the visibilities of $\kappa$~Tuc at six baselines. Using analytical solutions of a limb-darkened photosphere surrounded by uniform disc emission, we were able to derive significant (${>}3\sigma$) dust-to-star flux ratios of $\unit[5-7]{\%}$ in the wavelength range $\unit[3.37]{\mu m}-\unit[3.85]{\mu m}$.  Since the measured closure phases are close to zero, the \mbox{MATISSE} data strongly support the scenario that the excess is caused by circumstellar dust emission and not by a companion. The results present a further confirmation of the existence of hot exozodiacal dust around $\kappa$~tuc that has been detected previously (\citealt{Ertel2014,Ertel2016}). In particular, the existence confirms that the temporal variability seen in previous PIONIER observations is caused by a variability of the dust properties. An explanation for the origin of the variability is beyond the scope of our study.
 
 The newly derived \textit{L}~band fluxes and the previously published \textit{H}~band fluxes provided the basis for a SED modelling. The best-fits were obtained for amorphous carbon grains of size $a=\unit[0.58]{\mu m}$ though other carbons and astronomical silicate cannot be ruled out. Moreover, the grain  size cannot be confined within the $1\sigma$ confidence level. We note that the theoretical blow-out size of carbon grains around an F6 star amounts to $a_\text{BO}{\sim}\unit[1.4]{\mu m}$ (\citealt{Kirchschlager2013}) and smaller grains should be blown out of the system. However, this value neglects additional trapping mechanisms and has to be considered with caution (see e.g.~\citealt{Kral2017} for discussion).
  
 Since the \textit{H}~band data revealed a temporal variability, we combined them with the \textit{L}~band data in different ways. Depending on the approach, the best fits are obtained for a narrow dust ring at a stellar distance in the $\unit[0.1-0.29]{au}$\break range, and thus with$\,$a$\,$temperature$\,$between $\unit[940]{K}\,$and$\,\unit[1430]{K}$\break and  total dust mass between $\unit[2\times10^{-9}]{M_\oplus}$ and $\unit[10.4\times10^{-9}]{M_\oplus}$. Within the $1\sigma$ confidence level dust location and temperature are constrained to $\unit[0.032-1.18]{au}$ and $\unit[600-2000]{K}$. 
 
 The MATISSE observations open a new window to study hot exozodis. Though the \textit{L}~band data alone can hardly determine the dust properties, the combination of \textit{L}~band and NIR~data (\textit{H}~band) constrain the stellar distance of the emission. For a better understanding of the dust properties, and in particular for the location of the hot exozodi emission, simultaneous observations of $\kappa$~Tuc in the NIR and MIR domain are required in the future. The NIR flux has varied with a period of ${\sim}12$ months between 2012 and 2014. Therefore, \mbox{PIONIER} and \mbox{MATISSE} observations conducted within a few months will allow to further constrain the dust location, and several sequences of these combined observations will help to understand the flux variability. Observations with VLTI/GRAVITY in \textit{K}~band could support the study of the hot exozodi by filling the gap in the SED at an intermediate wavelength. The detection or non-detection of dust emission in \textit{N}~band using MATISSE observations of higher quality can potentially determine or rule out the presence of silicate material due to the prominent silicate feature at $\lambda\unit[{\sim}10]{\mu m}$. Finally, high spectral resolution observations by MATISSE can be used to trace \textit{LM}~band features that are consistent with certain materials. We note that $\kappa$~Tuc has a declination of $\delta\approx-69^\circ$ which is too far south to be visible for CHARA/FLUOR (\textit{K}~band) or the Large Binocular Telescope Interferometer (\textit{N}~band).
%
 
 We have proven in this study that MATISSE offers the required sensitivity and spatial resolution for the observation of hot exozodis. In the future, the \textit{L}~band and upcoming \textit{M}~band observations will allow to determine the properties of other hot exozodis. Given the dust-to-star flux ratio of up to $\unit[7]{\%}$ in \textit{L}~band is a frequent phenomenon, MATISSE will most likely allow the discovery of new hot exozodis.\vspace*{-0.7cm}

\section*{Acknowledgements}
\vspace*{-0.2cm}Based on observations collected at the European Organisation for Astronomical Research in the Southern Hemisphere under ESO ID 0103.C-0725(A). FK was supported by European Research Council Grant SNDUST ERC-2015-AdG-694520. SW and AVK acknowledge support by the DFG through grants WO~857/15-2 and KR~2164/15-2. This research has benefited from the help of SUV, the VLTI user support service of the Jean-Marie Mariotti Center ({\href{http://www.jmmc.fr/suv.htm}{http://www.jmmc.fr/suv.htm}).\\[-1.1cm]
 \section*{Data availability}
\vspace*{-0.2cm}The data underlying this article will be shared on reasonable request to the corresponding author. The raw data are publicly available at the ESO archive and at the optical interferometry database of the JMMC.
\\[-1.1cm]

  \bibliographystyle{mnras}
 \bibliography{Literature} 
\label{lastpage}


\end{document}